\documentclass[11pt]{article}
\usepackage{graphicx,rotating}
\usepackage{epsfig}
\usepackage{amssymb}
\usepackage{color}
\usepackage{colordvi}

\newcommand{\<}{\langle}
\renewcommand{\>}{\rangle}
\newcommand{\T}{\mathcal T}
\newcommand{\PP}{\mathcal P}
\newcommand{\C}{\mathcal C}
\newcommand{\rs}{\mbox{\bbold R}}
\newcommand{\M}{\mathcal M}

 \newfont{\bbold}{msbm10}

\newtheorem{Theorem}{Theorem}

\begin{document}

\title{Self-avoiding walks crossing a square}
\author{M. Bousquet-M\'elou$^1$, A. J. Guttmann$^2$ and I. Jensen$^2$\thanks{bousquet@labri.fr, tonyg,I.Jensen@ms.unimelb.edu.au} \\
$^1$ CNRS, LaBRI, Universit\'e  Bordeaux I, 351 cours de la Lib\'eration,\\
33405 Talence Cedex, France\\
$^2$ ARC Centre of Excellence for Mathematics and Statistics of Complex Systems, \\
Department of Mathematics and Statistics, \\
The University of Melbourne, Victoria 3010, Australia}

\date{\today}

\maketitle

\begin{abstract}
We study a restricted class of self-avoiding walks (SAW) which start at the origin (0, 0),
end at $(L, L)$, and are entirely contained in the square $[0, L] \times [0, L]$ on the
square lattice ${\mathbb Z}^2$. The number of distinct walks is known to grow as $\lambda^{L^2+o(L^2)}$.
We estimate $\lambda = 1.744550 \pm 0.000005$ as well as obtaining strict upper and lower bounds,
$1.628 < \lambda < 1.782.$
We give exact results for the number of SAW of length $2L + 2K$ for $K = 0, 1, 2$ and asymptotic
results for $K = o(L^{1/3})$.

We also consider the model in which a weight or {\em fugacity} $x$ is
associated with each step of the walk. This gives rise to a canonical model of a phase transition.
For $x < 1/\mu$ the average length of a SAW grows as $L$, while for $x > 1/\mu$ it  grows as
 $L^2$. Here $\mu$ is the growth constant of unconstrained SAW in ${\mathbb Z}^2$.
For $x = 1/\mu$ we provide numerical evidence, but no proof, that the average walk length
grows as $L^{4/3}$. Another problem we study is that of SAW, as described above, that pass
through the central vertex of the square. We estimate the proportion of such walks as a fraction
of the total, and find it to be just below $80\%$ of the total number of SAW.

We also consider Hamiltonian walks under the same restriction. They are known to grow as
$\tau^{L^2+o(L^2)}$ on the same $L \times L$ lattice. We give precise estimates for $\tau$ as
well as upper and lower bounds, and prove that $\tau < \lambda.$
\end{abstract}

\section{Introduction \label{sec:intro}}

We consider the problem of self-avoiding walks on the square lattice ${\mathbb Z}^2.$
For walks on an infinite lattice, it is generally accepted \cite{MS} that the number of such
walks of length $n$, equivalent up to a translation, denoted $c_n$, grows as
$c_n \sim const. \mu^n n^{\gamma-1},$ with metric properties, such as mean-square radius of gyration 
or mean-square end-to-end distance growing as $\<R^2\>_n \sim const. n^{2\nu},$ where $\gamma = 43/32$ 
and $\nu = 3/4.$ The growth constant $\mu$ is lattice dependent, and for the square lattice is not 
known exactly, but is indistinguishable numerically from the unique positive root of the equation 
$13x^4 - 7x^2 - 581 = 0.$ We denote the generating function by $C(x) := \sum_n c_n x^n.$ It will be 
useful to define a second generating function for those SAW which start at the origin $(0,0)$ and 
end at a given point $(u,v),$ as $G_{(0,0;u,v)}(x).$ In terms of this generating function, the 
{\em mass} $m(x)$ is defined \cite{MS} to be the rate of decay of $G$ along a coordinate axis,
\begin{equation}\label{eq0}
m(x) := \lim_{n \to \infty} \frac{-\log G_{(0,0;n,0)}(x)}{n}.
\end{equation}

Here, we are interested in a restricted class of square lattice SAW which start at the origin 
$(0,0)$, end at $(L,L),$ and are entirely contained in the square $[0,L] \times [0,L].$ 
A {\em fugacity}, or weight,  $x$ is associated with each step of the walk. Historically, this
problem seems to have led two largely independent lives. One as a problem in combinatorics
(in which case the fugacity has been implicitly set to $x=1$), and one in the statistical mechanics
literature where the behaviour as a function of fugacity $x$ has been of considerable interest, as
there is a fugacity dependent phase transition.

The problem seems to have first been seriously studied as a mathematical problem
 by Abbott and Hanson \cite{AH} in 1978,
many of whose results and methods are still powerful today. A key question considered
both then and now, is the number of distinct
SAW on the constrained lattice, and their growth as a function of the size of the lattice.
Let $c_n(L)$ denote the number of $n$-step SAW which start at the origin $(0,0)$,
end at $(L,L)$ and are entirely contained in the square $[0,L] \times [0,L].$
Further, let $C_L(x) := \sum_n c_n(L)x^n.$ Then $C_L(1)$ is the number of distinct walks
from the origin to the diagonally opposite corner of an $L \times L$ lattice. In \cite{AH},
and independently in \cite{WG}, it was proved that
$C_L(1)^{1/L^2} \rightarrow \lambda.$
The value of $\lambda$ is not known, though bounds and estimates have been given in \cite{AH,WG}.
One of our purposes in this paper is to improve on both the bounds and the estimate.

Like so many problems in lattice statistics, this one owes a debt to J. M. Hammersley. A
closely related problem to the one considered here is discussed in \cite{OW79}, which is
in turn devoted to problems posed by Hammersley. However the earliest mention of this problem
appears to be by Knuth \cite{K76}, who calculated the number of SAW crossing a $10 \times 10$
square by Monte Carlo methods, and estimated the number to be $(1.6 \pm 0.3) \times 10^{24}.$
It is now known, see Table~\ref{tab:totser} below, that the correct answer is $1.5687.. \times 10^{24}.$
A related problem was studied by Edwards  in \cite{E85}.  He considered SAW
 starting at a point denoted the origin with end point
a distance $L$ from the origin, and no other points at distance $L$ or greater. Let $g(L)$
denote the number of such SAW. Then Edwards proved that $\lim_{L \to \infty} g(L)^{(1/L^2)}$
exists and lies between 2.3 and 5.0. In our notation, Edwards has proved that $1.53 < \lambda < 2.24.$
Edwards also proved that the same limit holds for SAW from the origin to the boundary of any
convex, bounded subset of ${\mathbb Z}^2.$ His numerical work led him to suggest that
$\lambda$ is about 1.77. Our best estimate, given below, is 1.744550(5).

The problem of Hamiltonian paths on an $L \times M$ rectangular grid, going from $(0,0)$ to $(L,M)$ 
has also been considered previously. Earlier work is described in \cite{CK97}, where Collins and 
Krompart also give generating functions for the number of such paths on grids with $M = 1,2,3,4,5.$
In \cite{J98} Jacobsen and Kondev gave a field-theoretical estimate of the growth constant 
for Hamiltonian SAW on the square lattice, which must fill a square, as $1.472801 \pm 0.00001$.

In the statistical mechanics literature, the problem appears to have been introduced by
Whittington and Guttmann \cite{WG} in 1990, who were particularly
interested in the phase transition that takes place as one varies the fugacity associated
with the walk length. All walks on lattices up to  $6 \times 6$ were enumerated, and the
estimate $\lambda = 1.756 \pm 0.01$ was given. At a critical value, $x_c$ the average
walk length of a path on an
$L \times L$ lattice changes from $\Theta(L)$ to $\Theta(L^2),$  where we define $\Theta(x)$ as follows:
Let  $a(x)$ and $b(x)$ be  two functions of some variable $x$. We
write  that $a(x) = \Theta( b(x))$ as $x\rightarrow x_0$ if there
exist two positive constants $\kappa_1$ and $\kappa_2$ such that,
for $x$ sufficiently close to $x_0$,
$$
\kappa_1\; b(x) \le a(x) \le \kappa_2\; b(x).
$$
In \cite{WG} the critical fugacity was proved to be at least $1/\mu$,
its value was estimated numerically and was conjectured to be
$x_c = 1/\mu$, and in \cite{M} the conjecture was proved by Madras.

The problem was subsequently taken up by Burkhardt and Guim \cite{BG91}, who extended the enumerations
given in \cite{WG} to $9 \times 9$ lattices, and used their data to give the improved estimate 
$\lambda = 1.743 \pm 0.005.$ By considering SAW as the $N \to 0$ limit of the O$(N)$ model of magnetism,
Burkhardt and Guim show that the conjecture $x_c = 1/\mu$ made in \cite{WG} on numerical
grounds follows directly, though this is not a proof, unlike the subsequent result of Madras \cite{M}.

They also gave a scaling {\em Ansatz}
for the behaviour of $C_L(x)$ for $L$ large in the vicinity of $x=x_c.$ They proposed
\begin{equation}\label{BG}
C_L(x) \sim L^{-\eta_c}f[L^{1/\nu}(x_c - x)]
\end{equation}

where $\nu = 3/4$, as described above,
and $\eta_c = 5/2$ is the corner exponent of the magnetisation \cite{C84}, given by
Cardy's \cite{C84} result $\eta_c(\theta) = \frac{\pi}{\theta}\eta_{\|},$ for a wedge-angle
$\theta$, which is $\pi/2$ in this case. $\eta_{\|} = 5/4$ is the surface exponent that characterises the
decay of spin-spin correlations parallel to the boundary in the semi-infinite geometry,
corresponding to wedge-angle $\pi.$ Consequences of this scaling Ansatz include the
following predictions:
$$C_L(x_c) \sim const. L^{-\eta_c}$$
\begin{equation}\label{DS}
\< n(x_c,L) \> = x\frac{\partial}{\partial x} C_L(x_c) \sim const. L^{1/\nu}
\end{equation}
$$\< (n(x_c,L) - \<n(x_c,L)\>)^2\> = (x\frac{\partial}{\partial x})^2 \ln C_L(x_c) \sim const. L^{2/\nu}.$$
They tested these results from their numerical data, and found them well supported. We
provide even firmer support for these results on the basis of radically extended numerical
data. Equation (\ref{DS}) has also previously been given by Duplantier and Saleur \cite{DS87}.

Burkhardt and Guim also considered a generalisation of the problem considered here by including a 
second fugacity, associated with steps in the boundary. This allows the problem of adsorbing 
boundaries to be studied. We will not discuss this aspect of the problem further, except to note 
that in \cite{BG91} a full scaling theory is developed, and the predictions of the theory are 
tested against numerical data.

In \cite{AH} the slightly more general problem of SAW constrained to an $L \times M$ lattice
was considered, where the analogous question was asked: how many non-self-intersecting
paths are there from $(0,0)$ to $(L,M)?$ If one denotes the number of such paths by
$C_{L,M},$ it is clear that, for $M$ finite, the paths can be generated by a finite
dimensional transfer matrix, and hence that the generating function is rational \cite{stanley-vol2}. 
Indeed, in \cite{AH} it was proved that
\begin{equation}\label{G2}
G_2(z) = \sum_{L \ge 0} C_{L,2} z^L = \frac{1-z^2}{1 - 4z + 3z^2 - 2z^3 -z^4},
\end{equation}
(where here we have corrected a typographical error). It follows that
$C_{L,2} \sim {\rm const.} \lambda_2^{2L},$ where
$\lambda_2 = \sqrt{\frac{2}{\sqrt{13}-3} }= 1.81735\ldots$.

\begin{figure}
\includegraphics[scale=0.9]{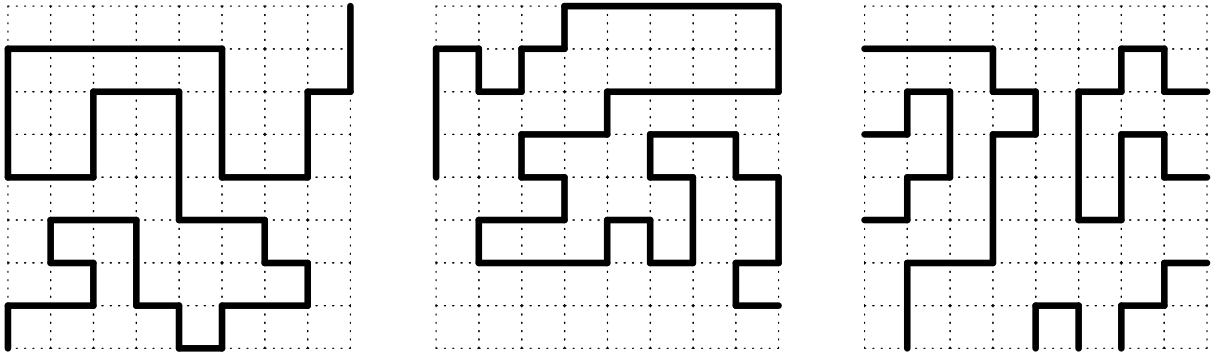}
\caption{\label{fig:example}
An example of a SAW configuration crossing a square (left panel), traversing
a square from left to right (middle panel) and a cow-patch (right panel).
}
\end{figure}

In this paper we also consider two further problems which can be seen as generalisations
of the stated problem. Firstly, we consider the problem where SAWs are allowed to start
anywhere on the left edge of the square and terminate anywhere on the right edge; so these are
walks {\em traversing} the square from left to right.
We call such walks {\em transverse\/} walks.
Secondly, we consider the problem
in which there may be several independent SAW, each SAW starting and ending on the
perimeter of the square.
The SAW are not allowed to take steps along the edges of the perimeter.
Such walks partition the square
 into distinct regions and by colouring the regions alternately black and white
we get a {\em cow-patch} pattern.
Each problem is illustrated in Figure~\ref{fig:example}.

Following the work in \cite{WG}, Madras  \cite{M} proved a number of theorems. In fact,
most of Madras's results were proved for the more general $d$-dimensional hyper-cubic lattice,
but here we will quote them in the more restricted two-dimensional setting.

\begin{Theorem}
The following limits,
$$
\mu_1(x) := \lim_{L \to \infty} C_L(x)^{1/L} \quad \hbox{ and }
\quad
 \mu_2(x) := \lim_{L \to \infty} C_L(x)^{1/L^2},
$$
are well-defined in $\rs \cup \{+\infty\}$.

More precisely,
\begin{itemize}
\item [$(i)$] $\mu_1(x)$
is finite for $0 < x \le 1/\mu$, and is infinite for
$x > 1/\mu.$ Moreover, $0 < \mu_1(x) < 1$ for $0 < x < 1/\mu$ and
$\mu_1(1/\mu) = 1.$
\item [$(ii)$]
 $\mu_2(x)$ is finite for all $x > 0.$
Moreover, $\mu_2(x) = 1$ for $0 < x \le 1/\mu$ and $\mu_2(x) >
1$ for $x > 1/\mu.$
\end{itemize}
\end{Theorem}

In \cite{WG} the existence of the limit $\mu_2(x)$ was proved, and in addition
upper and lower bounds on $\mu_2(x)$ were established.

The average length of a (weighted) walk is defined to be
\begin{equation}\label{eq3}
\<n(x,L)\> := \sum_n n c_n(L)x^n/\sum_n  c_n(L)x^n.
    \end{equation}

\begin{Theorem}
 For $0 < x < 1/\mu$, we have that $\<n(x,L)\>= \Theta(L)$ as
 $L\rightarrow  \infty$, while  for
 $x > 1/\mu$, we have
$\<n(x,L)\>=\Theta(L^2)$.
\end{Theorem}

In \cite{WG} it was proved that $\<n(1)\>_L=\Theta(L^2)$.
The situation at $x = 1/\mu$ is unknown. We provide
compelling numerical evidence that in fact $\<n(1/\mu)\>_L
=\Theta(L^{1/\nu})$ , where $\nu = 3/4$, in
accordance with an intuitive suggestion in both \cite{BG91} and \cite{M}.

\begin{Theorem}
For $x > 0$, define $f_1(x) = \log \mu_1(x)$ and
 $f_2(x) = \log \mu_2(x).$
\begin{itemize}
\item [$(i)$]
 The function $f_1$ is a strictly increasing, negative-valued convex
 function of
$\log x$ for $0 < x < 1/\mu,$ and $f_1(x)=\Theta ( -m(x))$ as $x \to
 1/\mu^-$, where $m(x)$ is the {\em mass}, defined by {\rm(\ref{eq0})}.
\item [$(ii)$]
The function $f_2$ is a strictly increasing, convex function of
$\log x$ for $x > 1/\mu,$ and satisfies $0 < f_2(x) \le \log \mu +
\log x.$
\end{itemize}
\end{Theorem}

Some, but not all of the above results were previously proved in \cite{WG}, but these three theorems
elegantly capture all that is rigorously known.

The rest of the paper is organised as follows: In the next section 
we describe our enumeration methods, and explain how they are used to obtain radically
extended series expansions for the number of walks crossing a square, the number of cow-patch
configurations and the number of transverse SAW. Section 3 details the results we have
obtained. In Section 4
we derive methods for obtaining rigorous upper and lower bounds on $\lambda.$ In that section we show
that upper bounds based on counting cow-patch configurations are fully equivalent to
the method of Abbott and Hanson, based on 0--1 admissible matrices. An improved method of
lower bounds based on counting transverse walks is also derived. In section 5 we then apply
these methods to our radically extended enumerations 
to provide significantly improved bounds on $\lambda.$ In section 6 we give exact results for
short SAW crossing a square. The shortest SAW that can cross a square from $(0,0)$ to $(L,L)$
is of length $2L.$ We give the exact number of such SAW of length $2L+2K,$ for $K = 0,1,2,$ and
asymptotic results for $K = {\rm o}(L^{1/3})$. Section 7 is devoted to a numerical 
analysis which gives precise (though non-rigorous) estimates of $\lambda,$ for all three
types of configurations, a discussion of the mean number of steps as a function of fugacity,
fluctuations in this quantity, and a scaling theory for such fluctuations. We also speculate
on the nature of the sub-dominant behaviour of the asymptotic form for the number of SAW.
Section 8 is also a numerical study, but of the number of SAW that pass through the central vertex of
an $L \times L$ square. Finally in section 9 we study Hamiltonian paths, obtaining both 
rigorous upper and lower bounds on the growth constant, and a numerical estimate.

\section{Exact enumeration \label{sec:enum}}

In the following we give a fairly detailed description of the algorithm
we use to enumerate the number of walks crossing a square and briefly
outline how this basic algorithm is modified in order to include a step fugacity,
study SAWs traversing a square and the cow-patch configurations.

\subsection{The basic algorithm}

We use a transfer matrix algorithm to count the number of walks crossing $L \times M$ rectangles.
The algorithm is based on the method of Conway et al. \cite{CEG} for enumerating ordinary
self-avoiding walks. The transfer matrix technique
involves drawing a boundary line through the rectangle intersecting
$M+1$ or $M+2$ edges.

\begin{figure}
\includegraphics[scale=0.9]{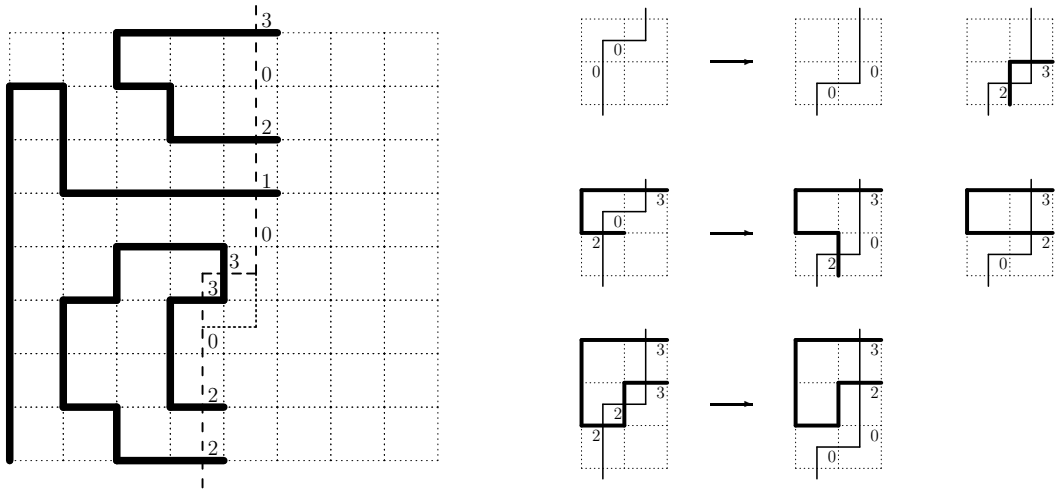}
\caption{\label{fig:transfer}
The left panel shows a snapshot of the intersection (dashed line) during the
transfer matrix calculation. Walks are enumerated by successive moves of the
kink in the boundary, as exemplified by the position given by the dotted line,
so that the $L \times M$ rectangle is built up one vertex at a time. To the left of the
boundary we have drawn an example of a partially completed walk. Numbers along
the boundary indicate the encoding of this particular configuration.
The right panel shows some of the local configurations which occur as the kink
in the intersection is moved one step.}
\end{figure}

For each configuration of occupied or empty edges we maintain a count
of partially completed walks intersecting the boundary in that pattern. Walks in
rectangles are counted by moving the boundary adding one vertex at a time
(see Figure~\ref{fig:transfer}). Rectangles are built up column by column with each
column constructed one vertex at a time. Configurations are represented by lists of
states $\{\sigma_i\}$,  where the value of the state $\sigma_i$ at position $i$ must
indicate if the edge is occupied or empty. An empty edge is indicated by $\sigma_i=0$.
An occupied edge is either free (not connected to other edges) or connected to exactly
one other edge via a path to the left of the boundary. We indicate this by $\sigma_i=1$
for a free end, $\sigma_i=2$ for the lower end of a loop and $\sigma_i=3$ for the upper
end of a loop connecting two edges. Since we are studying self-avoiding walks on a
two-dimensional lattice the compact encoding given above uniquely specifies which ends
are paired. Read from the bottom the configuration along the intersection in
Figure~\ref{fig:transfer} is $\{2203301203\}$ (prior to the move) and
$\{2300001203\}$ (after the move).

There are major restrictions on the possible configurations and their updating rules.
Firstly, since the walk has to cross the rectangle there is exactly one free end in
any configuration. Secondly, all remaining occupied edges are connected by a path to
the left of the intersection and we cannot close a loop. It is therefore clear that
the total number of 2's  equals the total number of 3's. Furthermore, as we look through
the configuration from the bottom the number of 2's is never smaller than the number of
3's (they are perfectly balanced parentheses).
We also have to ensure that the graphs we construct have only one connected component.
In the following we shall briefly show how this is achieved.

\begin{table}
\caption{\label{tab:update}
The various `input' states and the `output' states which arise as the
boundary line is moved in order to include one more vertex.
Each panel contains up to three possible `output' states or other allowed
actions.}
\begin{center}
\renewcommand{\arraystretch}{1.2}
\begin{tabular}{|c|ccc|ccc|cc|cc|}  \hline  \hline
\raisebox{-1.5mm}{Bottom}\raisebox{-0.5mm}{\Large $\backslash$}\raisebox{0.5mm}{Top}
 &\multicolumn{3}{c|}{0} &\multicolumn{3}{c|}{1} & \multicolumn{2}{c|}{2}
& \multicolumn{2}{c|}{3}  \\
\hline
0  &  $00$   & $23$  &
   &  $01$   & $10$ & Res
   &  $02$   & $20$
   &  $03$   & $30$    \\    \hline
1  & $01$   & $10$ & Res
      &  & &
   &  \multicolumn{2}{c|}{$\widehat{00}$}
   &  \multicolumn{2}{c|}{$\widehat{00}$}  \\  \hline
2  &  $02$   & $20$ &
   &  & $\widehat{00}$ &
   &  \multicolumn{2}{c|}{$\overline{00}$} & &
    \\  \hline
3   & $03$   & $30$ &
   &  & $\widehat{00}$ &
   &  \multicolumn{2}{c|}{$00$}
   &  \multicolumn{2}{c|}{$\overline{00}$}   \\
   \hline \hline
\end{tabular}
\end{center}
\end{table}

We call the configuration before and after the move the `source' and `target', respectively.
Initially we have just one configuration with a single `1' at position $0$ (all other
entries  `0') thus ensuring that we start in the bottom-left  corner. As the boundary line
is moved one step, we run through all the existing sources. Each source gives rise to one
or two targets and the count of the source is added to the count of the target
(the initial count of a target being zero). After a source has been processed it
can be discarded since it will make no further contribution.
Table~\ref{tab:update} lists the possible local `input' states and the `output'
states which arise as the kink in the boundary is propagated one step and the
various symbols will be explained below.

Firstly, the values of the 'Bottom' and
'Top' table entries refer to the edge-states of the kink prior to the move.
The Top (Bottom) entry is the state of the edge intersected by
the horizontal (lower vertical) part of the boundary.

Some of the updating rules are illustrated further in Figure~\ref{fig:transfer}.
The topmost panels represent the input state `00' having the allowed output states
`00' and `23' corresponding to leaving the edges empty or inserting a new loop,
respectively.
The middle panels represents the input state `20' with output states `20' and `02'
from the two ways of continuing the loop end (note that the loop has to be continued
since we would otherwise generate an additional free end not located at the allowed
positions in the corners).
The  bottommost panels represents the input state `22' as part of the configuration
$\{02233\}$. In this case we connect two loop ends and we thus join two separate
loops into a single larger loop. The matching upper end of the innermost
loop  becomes the new lower end of the joined loop. The relabeling of the
matching loop-end when connecting two `2's (or two `3's) is denoted by over-lining
in Table~\ref{tab:update}.
When we join loop ends to a free end (inputs `12', `21', `13', and '31') we have to
relabel the matching loop end as a free end. This type of relabeling is indicated by
the symbol $\widehat{00}$.
The input state `11' never occurs since there is only one free end.
The input state `23' is not allowed since connecting the two ends results in a closed loop
and we thus discard any configuration in which a closed loop is formed. It is quite easy to
avoid forming closed loops. We only have to be careful when the input is '03' or '30'.
If the upper end of the loop is continued along the vertical output edge we would form
a closed loop if the horizontal edge immediately below was a lower loop end, and we
just check the state of this edge and only proceed if it is not in state '2' (naturally
the upper loop end can always by continued along the horizontal output edge).

Finally, we have marked two outputs, from the inputs `01' and `10' with `Res',
indicating situations where we terminate free ends. This results in completed partial
walks and is only allowed if there are no other occupied edges in the source
(otherwise we would produce graphs with separate pieces) and if we are at the
top-most vertex (otherwise we would not cross the rectangle). The count for this
configuration is the number of walks crossing a rectangle of height $M$ and
length $L$ equal to the number of completed columns.

The time required to obtain the number of walks on $L\times M$ rectangles grows
exponentially with $M$ and linearly with $L$. Time and memory requirements are
basically proportional to the maximal number of distinct configurations along the
boundary line. When there is no kink in the intersection (a column has just been
completed) we can calculate this number, $N_{\rm conf}(M)$, exactly. Obviously the free end
cuts the boundary line configuration into two separate pieces. Each of these pieces
consists of `0's and an equal number of `2's and `3's with the latter forming
 a perfectly balanced parenthesis system.

Each piece thus corresponds to a Motzkin path~\cite[Ch.~6]{stanley-vol2}
(just map 0 to a horizontal step, 2 to a north-east step, and 3 to a south-east step).
The number of Motzkin paths $M_n$ with $n$ steps is easily derived
from the generating  function ${\mathcal M}(x) = \sum_{n} M_n x^n $,
which satisfies $\M= 1+x\M+x^2\M^2$, so that

\begin{equation}\label{eq:Motzkin}
{\cal M}(x) =  [1-x-\sqrt{(1+x)(1-3x)}]/2x^2.
\end{equation}
\noindent
The number of configurations $N_{\rm conf}(M)$ for a rectangle of height $M$
is simply obtained by inserting a
free end between two Motzkin paths,
so that the generating function $\sum_{M} N_{\rm conf}(M)  x^M$ is simply $x\M(x)^2$. The
Lagrange inversion formula gives
$$
 N_{\rm conf}(M)  = 2  \sum_{i \ge 0} \frac{(M+1)!}{i! (i+2)!
 (M-2i)!}.
$$
\noindent
When the boundary line has a kink the number of configurations exceeds $N_{\rm conf}(M)$
but clearly is less than $N_{\rm conf}(M+1)$. From (\ref{eq:Motzkin}) we see that asymptotically 
$N_{\rm conf}(M)$ grows like $3^M$ (up to a power of $M$). So the same is true for the maximal
number of boundary line configurations and hence for the computational complexity of the
algorithm. Note that the total number a walks grows like $\lambda^{LM},$ so  our algorithm
leads to a better than exponential improvement over direct enumeration.

The integers occurring in the expansion become very large so the calculation
was performed using modular arithmetic \cite{Knuth}. This involves performing
the calculation modulo various prime numbers $p_i$ and then reconstructing the
full integer coefficients at the end. We used primes of the form $p_i=2^{30}-r_i$
where $r_i$ are distinct integers, less than 1000, such that $p_i$
is a (different) prime for each value of $i$.
The Chinese remainder theorem ensures that any integer has a unique representation
in terms of residues. If the largest integer occurring in the final expansion is $m$,
then we have to use a number of primes $k$ such that $p_1p_2\cdots p_k > m$.

\subsection{Extensions of the algorithm}

The algorithm is easily generalised to include a step fugacity
$x$. The count associated with the boundary line configuration has
to be replaced by a generating function for partial walks. Since
we only use this generalisation to study walks crossing an
$L\times L$ square the generating function is just a polynomial of
degree (at most) $L(L+2)$ in $x$. The coefficient of $x^n$ is just
the number of partial walks of length $n$ intersecting the
boundary line in the pattern specified by the configuration. The
generating function of the source is multiplied by $x^m$ and added
to the target, where $m$ is the number of additional steps
inserted. Not all $L(L+2)$ terms in the polynomials need be
retained. Firstly, any walk crossing the square has even length.
Thus in the generating functions for partial walks either all the
even or all the odd terms are zero, and we need only retain the
non-zero terms. Secondly, in order to construct a given boundary
line configuration, a certain minimal number of steps $n_{\rm
min}$ are required. Terms in the generating function of degree
lower than  $n_{\rm min}$ are therefore zero and again we need not
store these.

The generalisation to traversing walks is also quite simple. Firstly, we have
$M+1$ initial configurations which are empty except for a free end at position
$0\leq j \leq M$. This corresponds to the $M+1$ possible starting positions for
the walk on the left boundary. Secondly, we have to change how we produce
the final counts. The easiest way to ensure that a walk spans the rectangle
and that only single component graphs are counted is as follows: When column
$L+1$ has been completed we look at the $M+1$ configurations with a single
free end and add the counts from all of them. This is the number of walks traversing
an $L\times M$ rectangle.

The generalisation to cow-patch patterns is more complicated. Graphs can now have
many separate components each of which is a SAW, and there can thus be many free
ends in a boundary line configuration. Note that each SAW starts and terminates with
a step perpendicular to the border of the rectangle and there are never any steps
along the edges of the borders of the rectangle. There are
$2^{M-1}$ initial configurations since any of the edges in the first column from
position 1 to $M-1$ can be occupied by a free end or be empty (recall that in
cow-patch configurations the top and bottom-most horizontal edges cannot be occupied).
There is an extra
updating rule in the bulk in that we can have the local input `11' (joining
of two free ends) with the only possible output being `00'. Also the updating rules
at the upper and lower borders of the rectangle are different in this case.
At the upper border we only have the input `00' with the outputs `00'
and `10'  corresponding to the insertion of a free end on a vertical edge at
the upper border. There is no `23' or `01' outputs since these would produce
an occupied edge along the upper border.
At the lower border we have inputs `00', `01', and `02' and in each case
the only possible output is `00' (with the appropriate relabeling in the `02'
case).
Finally, the count of the number of cow-patch patterns is obtained by
summing over all boundary line configurations after the completion
of a column.

\section{Results \label{sec:enumres}}

As discussed above, in order to obtain the exact value of the
number of SAW crossing a square, some of which
are integers with nearly 100 digits, we performed the enumerations several times, each
time {\em modulo} a different prime. The enumerations were then  reconstructed
using the Chinese Remainder Theorem. Each run for a $19 \times 19$ lattice took about 72 hours
using 8 processors of a multiprocessor 1 GHz Compaq Alpha computer. Ten such runs were
needed to uniquely specify the resultant numbers.

Proceeding as above, we have calculated $c_n(L)$ for all $n$ for $L
\le 17.$
In other words, we have obtained the polynomials $C_L(x)$ for $L\le
17$.
In addition, we have computed $C_{18}(1)$ and $C_{19}(1),$ the total number of SAW crossing
an $18 \times 18$ and $19 \times 19$ square respectively. We have also computed the
corresponding quantities for cow-patch and transverse SAWs, denoted
$P_L(1)$ and $T_L(1)$ respectively, for $L \le 19.$
These are given in Table~\ref{tab:totser}.

In \cite{AH} the question was asked whether $C_{L,M}^{\frac{1}{LM}}$ is decreasing in
both $L$ and $M$? We can answer this in the negative, based on our enumerations.

\begin{table}
\centering
\caption{\label{tab:totser} The total number of walks crossing a square, $C_L(1)$,
cow-patch walks, $P_L(1)$ and traversing walks, $T_L(1)$.}
\renewcommand{\arraystretch}{1.05}
\scriptsize
\begin{tabular}{rl}  \hline  \hline
  $L$  &  $C_L(1)$ \\
\hline
 1 & 2  \\
 2 & 12  \\
 3 & 184  \\
 4 & 8512  \\
 5 & 1262816  \\
 6 & 575780564  \\
 7 & 789360053252  \\
 8 & 3266598486981642  \\
 9 & 41044208702632496804  \\
10 & 1568758030464750013214100  \\
11 & 182413291514248049241470885236  \\
12 & 64528039343270018963357185158482118  \\
13 & 69450664761521361664274701548907358996488  \\
14 & 227449714676812739631826459327989863387613323440  \\
15 & 2266745568862672746374567396713098934866324885408319028  \\
16 & 68745445609149931587631563132489232824587945968099457285419306  \\
17 & 6344814611237963971310297540795524400449443986866480693646369387855336  \\
18 & 1782112840842065129893384946652325275167838065704767655931452474605826692782532  \\
19 & 1523344971704879993080742810319229690899454255323294555776029866737355060592877569255844  \\
\hline
  $L$  &  $\frac12 P_L(1)$ \\
\hline
  1  &  1 \\
  2  &  7 \\
  3  &  160 \\
  4  &  11408 \\
  5  &  2522191 \\
  6  &  1718769373 \\
  7  &  3598611604598 \\
  8  &  23098353998190640 \\
  9  &  453839082673896579243 \\
 10  &  27266319759961440667165921 \\
 11  &  5005013940387988257218110301496 \\
 12  &  2805250606288167736619664411164848668 \\
 13  &  4798636658841347169993094278185741344065154 \\
 14  &  25042563713780942969666110695844976426050692260400 \\
 15  &  398585071868378544875200967972920693215965420927547891443 \\
 16  &  19343509060397504009184634223201418820841655935064055180184148711 \\
 17  &  2861743739297615012905209591294651941414000218185488280077237678797763881 \\
 18  &  1290420684731131093964422300362403673911432011198730662653676329480448243238167005 \\
 19  &  1773260101104126884305729846781529391070539884533101171392023893295633931250883380602647575 \\
\hline
  $L$  &  $T_L(1)$ \\
\hline
 1 & 8  \\
 2 & 95  \\
 3 & 2320  \\
 4 & 154259  \\
 5 & 30549774  \\
 6 & 17777600753  \\
 7 & 30283708455564  \\
 8 & 152480475641255213  \\
 9 & 2287842813828061810244  \\
10 & 102744826737618542833764649  \\
11 & 13848270995235582268846758977770  \\
12 & 5613766870113075134552249300590982081  \\
13 & 6856324633418315229580098999727214234534626  \\
14 & 25264653780547704599613926971040640439380254497299  \\
15 & 281194924965510769640501069703642937039678809002355743600  \\
16 & 9461739046646537749639494171503923182753987897972167546351180871  \\
17 & 963236702020101408274810653629921860636656580683490560257709270360444788  \\
18 & 296872411379358777499142156584947972393781613934413706389772635139720532797697401  \\
19 & 277150300263332125727926989254635730407844207233646123561354535935393720183262709640734296  \\
\hline \hline
\end{tabular}
\end{table}

\section{Proofs of bounds \label{sec:altbounds}}

Let $\C(L)$ be the set of self-avoiding walks crossing the $L\times L$ square from
its south-west corner $(0,0)$ to its  north-east corner $(L,L)$. Let
$C(L)$ denote the cardinality of $\C(L)$.
Let $\T(L)$ be the set of self-avoiding walks that {\em traverse}, the $L\times
L$ square: by this, we mean that the walk starts from the west edge of
the square and ends on the east edge (Figure~\ref{fig:example}). Let $T(L)$ be
the cardinality of $\T(L)$. Finally, let $\PP(L)$ be the set of 
{\em cow-patches}, of size $L$: a {\em cow-patch} is a configuration of mutually
avoiding self-avoiding walks on the $L\times L$ square, such that each walk has both
endpoints on the border of the square, but never contains an edge of
the border (Figure~\ref{fig:example}). Let $P(L)$ be the number of
cow-patches of size $L$.

We first prove in this section that 
\begin{equation} \label{same-limit} 
\lim C(L)^{1/L^2}= \lim T(L)^{1/L^2}= \lim P(L)^{1/L^2}= \lambda. 
\end{equation}
Then, we prove the following bounds on $\lambda$: for $L\ge 1$,
$$
C(L)^{1/(L+1)^2} \le \lambda, \quad T(L)^{1/((L+1)(L+2))} \le \lambda,
\quad \lambda \le (2P(L))^{1/L^2}.
$$

\begin{figure}
\begin{center}
\includegraphics[scale=1]{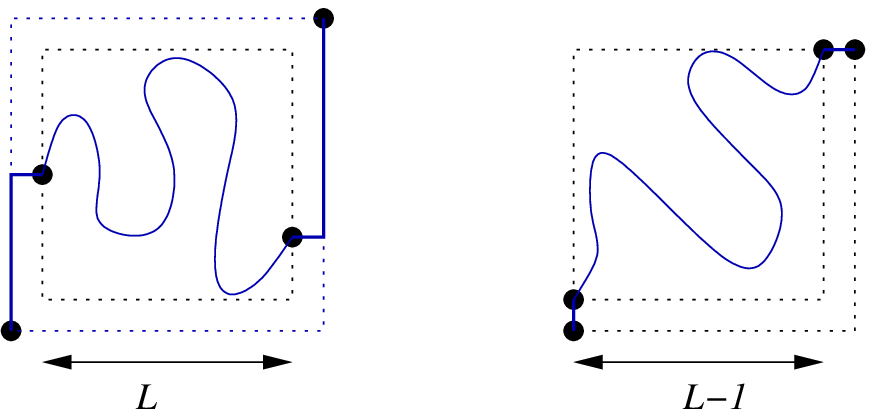}
\caption{\label{fig:simple-extensions}
From transverse walks to walks crossing a square (left).  From walks
crossing a square to cow-patches (right).}
\end{center}
\end{figure}

Let us first focus on~(\ref{same-limit}). As recalled in the previous
sections, the convergence of $C(L)^{1/L^2}$ to $\lambda$ has been
proved in earlier papers~\cite{AH,WG}. For walks of $\T(L)$, a similar
result follows from the fact that
$$
 C(L)\le T(L)\le C(L+2).
$$
The first inequality above is obvious. The second one is explained on
the left of Figure~\ref{fig:simple-extensions}.

\begin{figure}
\begin{center}
\includegraphics[scale=0.8]{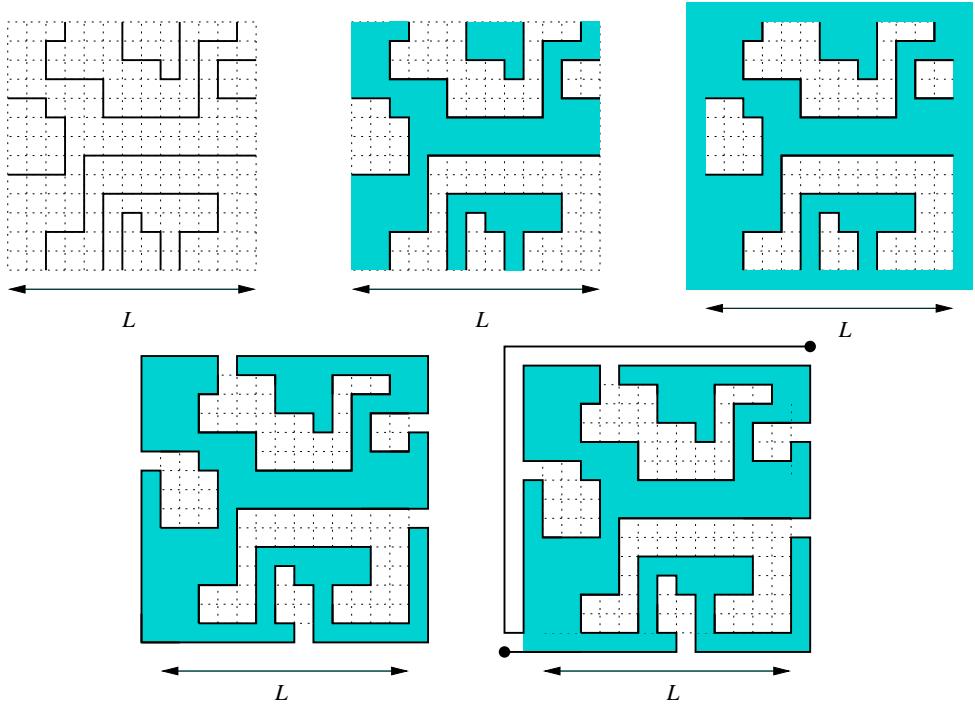}
\caption{\label{fig:tunnel}
From cow-patches to walks crossing a square.}
\end{center}
\end{figure}

For cow-patches, the existence and value of $\lim P(L)^{1/L^2}$
follows from
$$
C(L-1) \le P(L) \le C(L+3) .
$$
The first inequality is explained on the right of
Figure~\ref{fig:simple-extensions}. The second one is
a bit more tricky. We borrow the following argument from~\cite{E85}. It
is illustrated in Figure~\ref{fig:tunnel}.  Start from a cow-patch of size
$L$. Colour all cells of the square in black  and white, in
such a way that the south-west corner of the square is black and each step
included in one of  the walks of the cow-patch is adjacent to a black
cell and a white one. Surround the square by a layer of black cells,
so as to obtain a square of size $L+2$, containing a certain number of
white regions. For {\em each} white region, dig a tunnel (exactly one
tunnel) in the outer layer
to connect it to the outer world. In the figure thus
obtained, the border of the black region forms a self-avoiding 
{\em polygon}, that includes each walk of the cow-patch. It remains to
extend this polygon in a canonical way to obtain a walk of $\C(L+3)$,
illustrated in the last panel of Figure~\ref{fig:tunnel}.

\begin{figure}
\begin{center}
\includegraphics[scale=1]{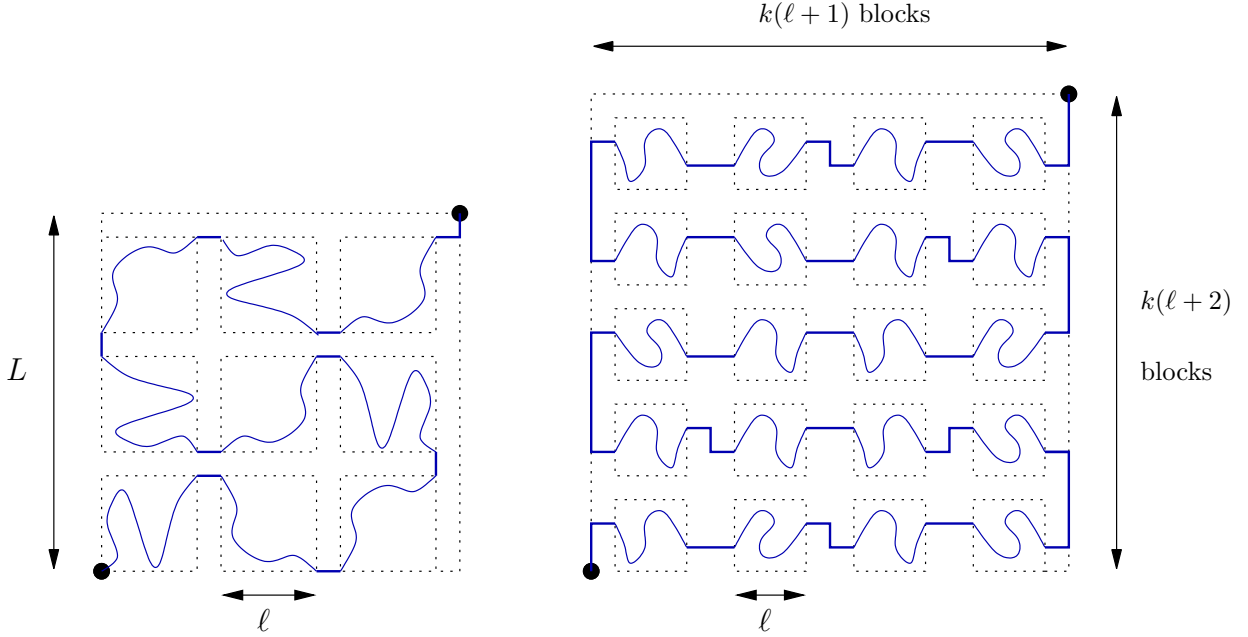}
\caption{\label{fig:c-bound}
Dense packings of walks crossing or traversing a square.}
\end{center}
\end{figure}

Let us now discuss lower and upper bounds on $\lambda$. The left-hand
side of Figure~\ref{fig:c-bound}
shows that for all $\ell$ and all odd $k$, it is possible to combine $k^2$
elements of $\C(\ell)$ to form an element of $\C(L)$ with
$L=k(\ell+1)$. In Figure~\ref{fig:c-bound}, $k=3$. This shows that
$$
C(\ell)^{k^2} \le C(L).
$$
Hence
$$
C(\ell)^{1/(\ell+1)^2} \le C(L)^{1/L^2}.
$$
Taking the limit as $k\rightarrow \infty $ implies that for all $\ell$,
$$
C(\ell)^{1/(\ell+1)^2} \le \lambda.
$$
Similarly, let us try to pack transverse walks densely. The
right-hand side of  Figure~\ref{fig:c-bound}
shows that for all $\ell$ and $k$, it is possible to combine $k^2(\ell+1)(\ell+2)$
elements of $\T(\ell)$ to form an element of $\C(L)$ with
$L=k(\ell+1)(\ell+2)$. This shows that
$$
T(\ell)^{k^2(\ell+1)(\ell+2)} \le C(L).
$$
Hence
$$
T(\ell)^{1/((\ell+1)(\ell+2))} \le C(L)^{1/L^2}.
$$
Taking the limit as $k\rightarrow \infty $ implies that for all $\ell$,
\begin{equation}\label{lower2}
\lambda \ge T(\ell)^{1/((\ell+1)(\ell+2))}.
\end{equation}

\begin{figure}
\begin{center}
\includegraphics[scale=1]{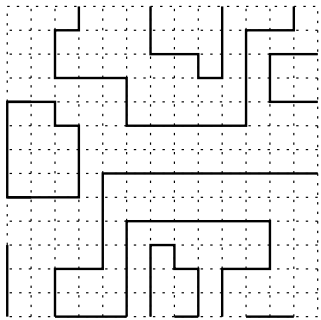}
\caption{\label{fig:generalized}
A generalized cow-patch.}
\end{center}
\end{figure}

Let us finally give upper bounds for $\lambda$. Define a coloured
cow-patch as a cow-patch in which the various regions are coloured in
black and white, in such a way that two adjacent regions have different
colours. Clearly, each cow-patch gives rise to 2 coloured cow-patches.
 Observe that there is
a bijection between coloured cow-patches of size $L$ and the
{\em admissible}, matrices of the same size, as defined in
Section~\ref{sec:bounds}. Since an element of $\C(L)$, with $L= k\ell$, can be seen
as the juxtaposition of $k^2$ admissible matrices (or coloured
cow-patches) of size $\ell$,
$$
C(L) \le (2P(\ell))^{k^2}.
$$
That is,
$$
C(L)^{1/L^2} \le (2P(\ell))^{1/\ell^2}
$$
and by letting $k\rightarrow \infty$, we obtain Abbott and Hanson's
bound: for all $\ell$,
$$
\lambda \le (2P(\ell))^{1/\ell^2}.
$$
One possible attempt to improve this bound is to consider {\em generalised}
cow-patches, in which the walks are allowed to include edges lying on
the west and south borders of the square
(Figure~\ref{fig:generalized}). Let $GP(L)$
denote the number of generalised cow-patches of size $L$. Since an
element of $\C(L)$, with $L= k\ell$, can be seen
as the juxtaposition of $k^2$ generalised patches, the above argument gives
$$
\lambda \le GP(\ell)^{1/\ell^2}.
$$
We have not exploited this improvement,
as it only changes
the fourth significant digit of our bound.

\section{Bounds
on the growth constant $\lambda$ \label{sec:bounds}}

For the more general problem of SAW going from $(0,0)$ to $(L,M)$ on
an $L \times M$ lattice, it was proved in \cite{AH} that
\begin{Theorem}
For each fixed $M$, $\lim_{L \to \infty} C_{L,M}^{\frac{1}{LM}} = \lambda_M$ exists.
\end{Theorem}
Further, Abbott and Hanson state that a similar proof can be used to establish  that 
$\lim_{L \to \infty} C_{L,L}^{\frac{1}{L^2}} := \lambda$ exists.
This was proved rather differently in \cite{WG}.

\subsection{Upper bounds on $\lambda$}
In \cite{AH} an upper bound on the growth constant $\lambda$ was
obtained by recasting
the problem in a matrix setting. We give below an alternative method
for establishing upper
bounds, based on defining a superset of paths. We then show that these
two methods are in fact identical.

Following \cite{AH}, consider any non-intersecting path
crossing the $L\times L$ square.
 Label each unit square in the $L \times L$ lattice by 1 if it lies to
 the right of the path, and by
0 if it lies to the left. This provides a
one-to-one
correspondence between paths and a subset
of $L \times L$ matrices with elements 0 or 1. Matrices corresponding to allowed paths
are called {\em admissible,} otherwise they are {\em inadmissible}. Since the total number
of $L \times L$ $0-1$ matrices is $2^{L^2},$ we immediately have the weak bound
$C_{L,L} \le 2^{L^2}.$ Of the 16 possible $2 \times 2$ matrices, only 14 can correspond to portions
of non-intersecting lattice paths. Note that there are only 12 actual paths from $(0,0)$ to $(2,2)$, 
but a further two matrices may correspond to paths that are embedded in a larger lattice.
Thus we find the bound $C_{L,L} \le 14^{(L/2)^2} ,$ so $\lambda \le 1.9343..$. Similarly,
for $3 \times 3$ lattices we find 320 admissible matrices (out of a possible 512), so
$\lambda \le  320^{1/9} = 1.8982..$ For $4 \times 4$ lattices, \cite{AH} claims that there
are 22662 admissible matrices, but we believe the correct number to be 22816, giving
the bound $\lambda \le 1.8723..$. We have made dramatic extensions of this work, using
a combination of finite-lattice methods and transfer matrices, as described below, and
have determined the number of admissible matrices up to $19 \times 19.$ There are
$3.5465202\ldots  \times 10^{90}$ such matrices, giving the bound
$$\lambda \le 1.7817.$$
This bound is fully equivalent to the bound $\lambda \le
(2P_L)^{1/L^2}$, where $P_L$ denotes
the number of cow-patch configurations
on the $L \times L$ lattice.
 This bound is proved below, in Section \ref{sec:altbounds},
 and the equivalence follows upon colouring cow-patches by two
 colours, such that adjacent
regions have different colours. Labeling the two colours $0$ and $1$
produces a $0-1$ matrix representation.

\subsection{Lower bounds on $\lambda$}
In \cite{AH} the useful bound
\begin{equation}\label{useful}
\lambda > \lambda_M^{\frac{M}{M+1}}
\end{equation}
 is proved.

The above evaluation of $\lambda_2,$  see (\ref{G2}),
immediately yields $\lambda > 1.4892\ldots$.

Based on exact enumeration, we have found the exact generating
functions
$G_M(z) = \sum_L C_{L,M} z^L$
for $M \le 6.$ For $M=3$ we find:
$$
G_3(z) =\frac{[1,-4, -4, 36, -39, -26, 50, 6, -15, 1]}
{[1,-12, 54, -124,133, 16, -175, 94, 69, -40, -12, 4, 1]},
$$
where we denote by $[a_0, a_1, \ldots, a_n]$ the polynomial $a_0+a_1z+
\cdots + a_nz^n$.
As explained above, all the generating functions $G_M(z)$ are
  rational.
  For $M = 4,5,6$, their numerator
and denominators are found to have degree $(26,27), (71,75)$ and
  $(186,186)$ respectively, in an obvious notation.

From these, we find the following values: $\lambda_3=1.76331\ldots$,
$\lambda_4=1.75146\ldots$,  $\lambda_5=1.74875\ldots$  and $\lambda_6=1.74728\ldots$.
Then from eqn. (\ref{useful}) and $\lambda_6$  we obtain the bound
$\lambda > 1.61339\ldots$.

However, an alternative lower bound can be obtained from transverse
SAWs, defined
in Section \ref{sec:intro}.
 If $T_L$
denotes the number of transverse SAW on the $L \times L$ lattice, then
we prove in the next section that

\begin{equation}\label{lower}
\lambda \ge T(L)^{1/((L+1)(L+2))}.
\end{equation}
 From our enumerations of $T(L)$, given above for $L \le 19,$ we obtain
 the improved  bound $\lambda > 1.6284.$

Combining our results for lower and upper bounds finally gives
$$1.6284 < \lambda < 1.7817.$$

\section{Short walks crossing a square \label{sec:small}}
As defined in the introduction,
let $c_n(L)$ be the number of $n$-step self-avoiding walks crossing an $L\times
L$ square. Clearly, this number is zero when  $n$ is
odd and also when $n<2L$. It is almost as
clear that
$$c_{2L}(L)={{2L} \choose L}.
$$
Indeed, there are $2L$
steps in the path, of which $L$ must go north and $L$ must go east.
Note that the number $c_{2L}(L)$ has asymptotic expansion
$$\frac{4^L}{\sqrt{L\pi}}\left(1 - \frac{1}{4L} + \frac{1}{128L^2} + \frac{5}{1024L^3} + \cdots\right).$$

\begin{figure}[hbt]
\begin{center}
\includegraphics[scale=1]{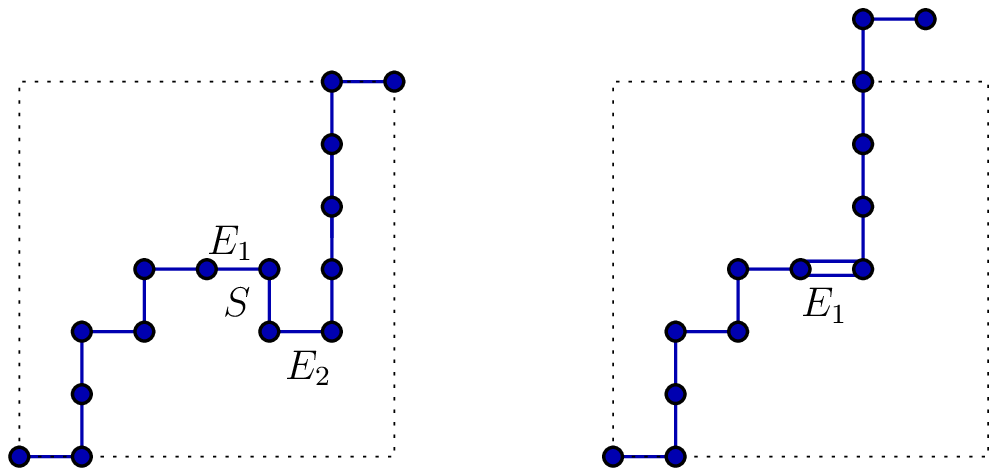}
\end{center}
\caption{Enumeration of self-avoiding walks with one vertical defect.}
\label{fig:one-defect}
\end{figure}

Let us now prove that
$$c_{2L+2}(L)=2L {2L \choose {L-2}}.$$  A walk
counted by $c_{2L+2}(L)$ has either $L+2$ vertical steps (and $L$
horizontal ones), or $L$ vertical steps (and $L+2$ horizontal ones). By
symmetry, we can focus on the first case. Let $w$ be such a
walk. We say that $w$ has a {\em vertical} defect., Among the $L+2$ vertical
steps of $w$, exactly one  goes south, while the $L+1$ others  go
north. The unique south step $S$ is necessarily preceded and followed
by an east
step, which we denote respectively $E_1$ and $E_2$. Let us mark $E_1$ and
delete $S$ and $E_2$ (Figure~\ref{fig:one-defect}). The marked path $w'$ thus obtained
allows one to recover
the original path $w$. It contains $L+1$ north steps and $L-1$ east
steps, one of which is marked. Moreover, the marked step cannot be at
ordinate $0$, nor at ordinate $L+1$.
Conversely, any walk  $w'$ satisfying these properties is obtained
(exactly once) from a walk counted by $c_{2L+2}(L)$ and having a
vertical defect.

The number of walks having  $L+1$ north steps and $L-1$ east
steps is ${{2L}\choose {L-1}}$. Marking one of the east steps gives a
factor $(L-1)$. Now we must subtract the number of walks in which the
marked step is either at level $0$ or at level $N+1$. Transforming the
marked step into a vertical step shows that each of these two families
of marked walks is in bijection with walks formed with $L+2$ up steps
and $L-2$ down steps. Putting these observations together gives
$$
c_{2L+2}(L)= 2 \left( (L-1) {{2L}\choose {L-1}} - 2 {{2L}\choose
  {L-2}}\right)
=2L  {{2L}\choose {L-2}}.
$$
Note that the number $c_{2L+2}(L)$ has the asymptotic expansion
$$\frac{L4^L}{\sqrt{L\pi}}\left(2 - \frac{33}{4L} + \frac{1345}{64L^2}  - \frac{23835}{512L^3} 
+  \cdots\right).$$

\begin{figure}[hbt]
\begin{center}
\includegraphics[scale=1]{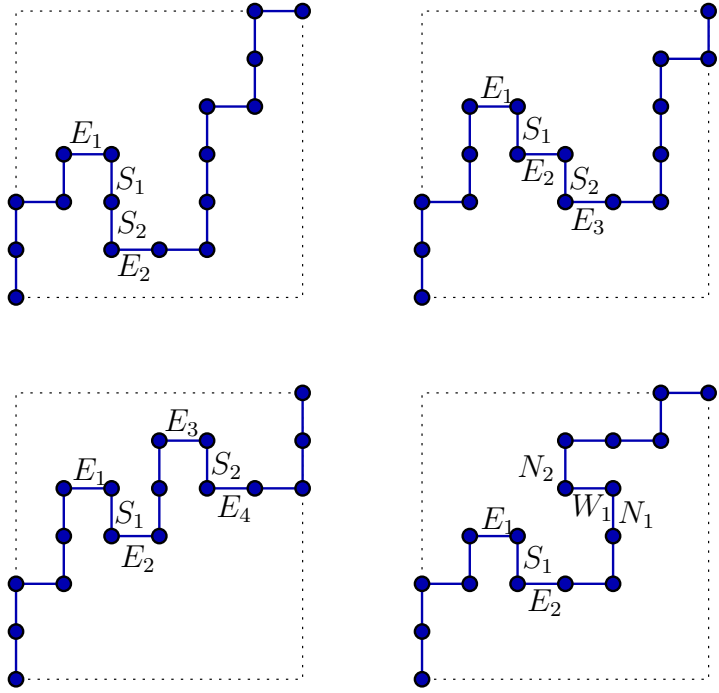}
\end{center}
\caption{Four types of self-avoiding walks with two defects.}
\label{fig:two-defects}
\end{figure}

The same ideas may be used to find the value of $c_{2L+4}(L)$. We will
prove that
\begin{equation}
\label{second-diag}
\frac 1 2 c_{2L+4}(L)= \frac{(2L)!}{L!(L+4)!}
\left( 48+90L+8L^2-28L^3-3L^4+4L^5+L^6\right) -2.
\end{equation}
First, note that $c_{2L+4}(L)/2$ is the number of self-avoiding walks (of length
$2L+4$, crossing the $L\times L$ square) in which the first defect,
that is, the first  {\em backward}, step, is a south step. We focus on
such walks, and study four distinct cases. The first three cases
count walks having two south steps, and the last case counts walks
having a south step
and a west step (Figure~\ref{fig:two-defects}).

\begin{enumerate}
\item The walk $w$ contains two adjacent south steps, $S_1$ and $S_2$. They
   are necessarily preceded by an east step $E_1$, and followed by
   another east step $E_2$. The walk has $L+4$ vertical steps and $L$
   horizontal steps. Mark $E_1$, and delete $S_1,S_2, E_2$ in
   order to obtain a walk $w'$ with $L+2$ north steps and $L-1$ east steps,
   one of which is marked. In $w'$, the marked step cannot be at level
   $0, 1, L+1$
   or $L+2$. Using the same ingredients as above, we obtain the number
   of such walks as
$$
(L-1){{2L+1}\choose {L-1}}-4{{2L+1}\choose {L-2}}.
$$
\item The walk contains a sequence $E_1S_1E_2S_2E_3$.  Again, $w$ has
  $L+4$ vertical steps and $L$
   horizontal steps. Mark $E_1$, and
  delete $S_1,E_2,S_2$ and $E_3$ in order to  obtain a walk with $L+2$ north
  steps and $L-2$ east steps,
   one of which is marked. In $w'$, the marked step cannot be at level $0, 1, L+1$
   or $L+2$. The number of such walks is
$$
(L-2){{2L}\choose {L-2}}-4{{2L}\choose {L-3}}.
$$
\item The walk contains a sequence $E_1S_1E_2$, and, further away, another
  sequence $E_3S_2E_4$, disjoint from the first one. Again, $w$ has
  $L+4$ vertical steps and $L$
   horizontal steps. Mark the steps
  $E_1$ and $E_3$, delete $S_1, E_2, S_2$ and $E_4$ in order to
  obtain a walk with $L+2$ north   steps and $L-2$ east steps,
   two of which are marked. Note that, in $w'$,   the first marked
  step cannot lie
  at level $0, L+1$ or $L+2$, while the second marked step cannot lie
  at level $0,1$ or $L+2$. Using the same ingredients as above,
  combined with the inclusion-exclusion principle, we find the number
  of such walks as
$$
{{L-2}\choose 2} {{2L}\choose {L-2}}-2 \left[
(L-3) {{2L}\choose {L-3}} - {{2L}\choose {L-4}}\right]
-4 {{2L}\choose {L-4}} - 2 {{2L}\choose {L-4}} + 5 {{2L}\choose
  {L-4}}
$$
$$
=
{{L-2}\choose 2} {{2L}\choose {L-2}}
-2(L-3) {{2L}\choose {L-3}} +{{2L}\choose {L-4}}.
$$
\item The walk $w$ contains a sequence $E_1S_1E_2$, and, further away, a
  sequence $N_1 W_1 N_2$ (with obvious notations). It thus contains
  $L+2$ vertical steps and $L+2$ horizontal ones. Mark the steps
  $E_1$ and $N_1$, delete $S_1, E_2, W_1$ and $N_2$ in order to
  obtain a walk $w'$ with $L$ north  steps and $L$ east steps,
  in which one step of each type is marked in such a way that the east
  marked step comes before the north marked step. In $w'$, the two
  marked steps
  cannot be consecutive (or $w$ would not be self-avoiding), the east
  marked step cannot lie at level $0$,
  and the north marked step cannot lie at abscissa $L$. Again, the
  inclusion-exclusion principle applies and gives the number of such
  walks as
$$
\frac 1 2 L^2 {{2L}\choose {L}} -
(2L-1){{2L-2}\choose {L-1}}
-2 L {{2L}\choose {L-1}}
+ 2 {{2L-1}\choose {L-1}}
+ \left[ {{2L}\choose {L}} -1\right] -1.
$$
\end{enumerate}
Putting together the four partial results we have obtained
gives~(\ref{second-diag}).
Note that the number $c_{2L+4}(L)$ has the  asymptotic expansion
$$\frac{L^24^L}{\sqrt{L\pi}}\left(2 - \frac{49}{4L} + \frac{2913}{64L^2}  - \frac{92971}{512L^3} 
+  \cdots\right).$$

The above argument suggests that it is very likely that, for every
fixed $K$, the sequence $c_{2L+2K}(L)$, for $L\ge 0$, is
{\em polynomially recursive}~\cite[Ch.~6]{S80,stanley-vol2}.

While it would probably be possible to find the number of possible paths of length $2L+6$, the number
of special cases that must be treated would become onerous.
We have therefore resorted
to a numerical study for walks of length $2L+2K,$  $K > 2$, based on our enumerations.
For $K = 3$ we found
$$\frac{L^3 4^L}{\sqrt{L\pi}}\left(\frac{4}{3} - \frac{49}{6L}   + \frac{1931 \pm 1}{64L^2} 
+ \cdots\right),$$
while the corresponding result for $K = 4$ is
$$\frac{L^4 4^L}{\sqrt{L\pi}}\left(\frac{2}{3}  + \frac{11}{4L} + \cdots\right).$$

We can give a heuristic argument
for the general form of the leading term in the
asymptotic expansion of the number of walks of length $2L+2K$
which gives as the leading order term
$\frac{4^L}{\sqrt{L\pi}}\frac{(2L)^K}{K!}.$ Here the first factor is given by the number
of ways of choosing the backbone, ${2L \choose L} \sim \frac{4^L}{\sqrt{L\pi}}$ and the
second is given by the number of ways of placing $K$ defects (or backward steps)
 on a path of length $2L$,
which is just $(2L)^K$. The defects are indistinguishable, introducing the factor $K!$.

This argument can be refined into a proof, for $K = {\rm o}(L^{1/3})$
by following the steps, {\em mutatis mutandis} in the proof of a similar result given
in \cite{EGRW}.

\section{Numerical analysis  \label{sec:ana}}

It has been proved \cite{AH,WG} that $\lim_{L\to \infty}C_{L,L}^{\frac{1}{L^2}}=\lambda$
exists. From this it is likely that $R_L = C_{L+1,L+1}/C_{L,L} \sim \lambda^{2L}$
though this has not been proved. Accepting this, the generating function ${\cal R}(x)= \sum_L R_Lx^L$
will have  radius of convergence
$x_c = 1/\lambda^2$, which we can estimate accurately using differential approximants
\cite{AJG}. In this way we estimate that for the crossing problem $x_c=0.32858(5)$, for the transverse
problem $x_c=0.3282(6)$ and for the cow-patch problem $x_c=0.328574(2)$. It is reassuring
to see, from our numerical studies,
that $\lambda$ appears to be the same for the three problems, as proved above, and we estimate
that $\lambda=1.744550(5)$.

We now speculate on the sub-dominant terms. For SAW on an infinite lattice, it is
widely accepted that $c_n \sim const.\mu^n n^g$
 where $c_n$ is the number of $n$ step SAW equivalent up to a translation.

It seems at least a plausible  speculation that, for SAW crossing an $L \times L$ lattice,
the number going from $(0,0)$ to $(L,L)$ is given by
$\sim  A\lambda^{L^2+bL}L^\alpha.$ We have investigated
this possibility numerically, and found it to be supported by the data, to some extent.

We fitted the data to the assumed form, fixing the value of $\lambda$ at our best estimate, 1.744550.
This then leaves two unknown parameters $b$ and $\alpha.$ For cow-patch walks we find $b \approx 0.8558$ 
and $\alpha \approx -0.500.$ This suggests asymptotic behaviour $A_P \lambda^{L^2+0.8558L}/\sqrt{L}$, 
and we estimate  $A_P \approx 0.52$. For transverse walks and walks crossing a square $b$ is quite small, 
most likely zero. A value of $b= 0$ would imply the absence of
a term O$(\lambda^{bL})$, or possibly the presence of a term O$(\log L)$, or some power of a logarithm.
We have investigated the latter possibility by including a logarithmic factor, and found that
the data does not support the presence of such a term for either class of walk. Of course,
we cannot rule out some small power of a logarithm, but this seems less likely than the
absence of a term O$(\lambda^{bL})$.

We next investigated the possibility that the subdominant term is O$(L^\alpha).$ A simple
ratio analysis \cite{AJG} then led to the estimates $\alpha = -0.7$ for walks crossing a
square, and $\alpha = 1.0$ for transverse walks. If our assumed form is correct, we expect
these estimates to be accurate to within 10-15\%. We also studied the sequence whose terms
are given by the quotient $T_L/C_L.$ This has the advantage that the $\lambda$ dependence
cancels, and so our result is independent of any uncertainty in the value of $\lambda.$ We
find that $T_L/C_L \sim const. L^{1.7}$ This is in agreement with the estimates of $\alpha$
found separately, for the two series. Thus we very tentatively speculate
that $C_L \sim 8\lambda^{L^2}/L^{0.7}$ and $T_L \sim 9\lambda^{L^2}L,$ where the amplitude estimates
follow by the simple expedient of fitting the assumed $L$ dependent form to the data, term-by-term,
and extrapolating the resulting sequence of amplitude estimates. Given the sensitivity of the
amplitudes to both $\lambda$ and $\alpha$, we do not feel confident quoting an uncertainty
for the amplitudes.

\begin{figure}
\includegraphics[scale=0.8]{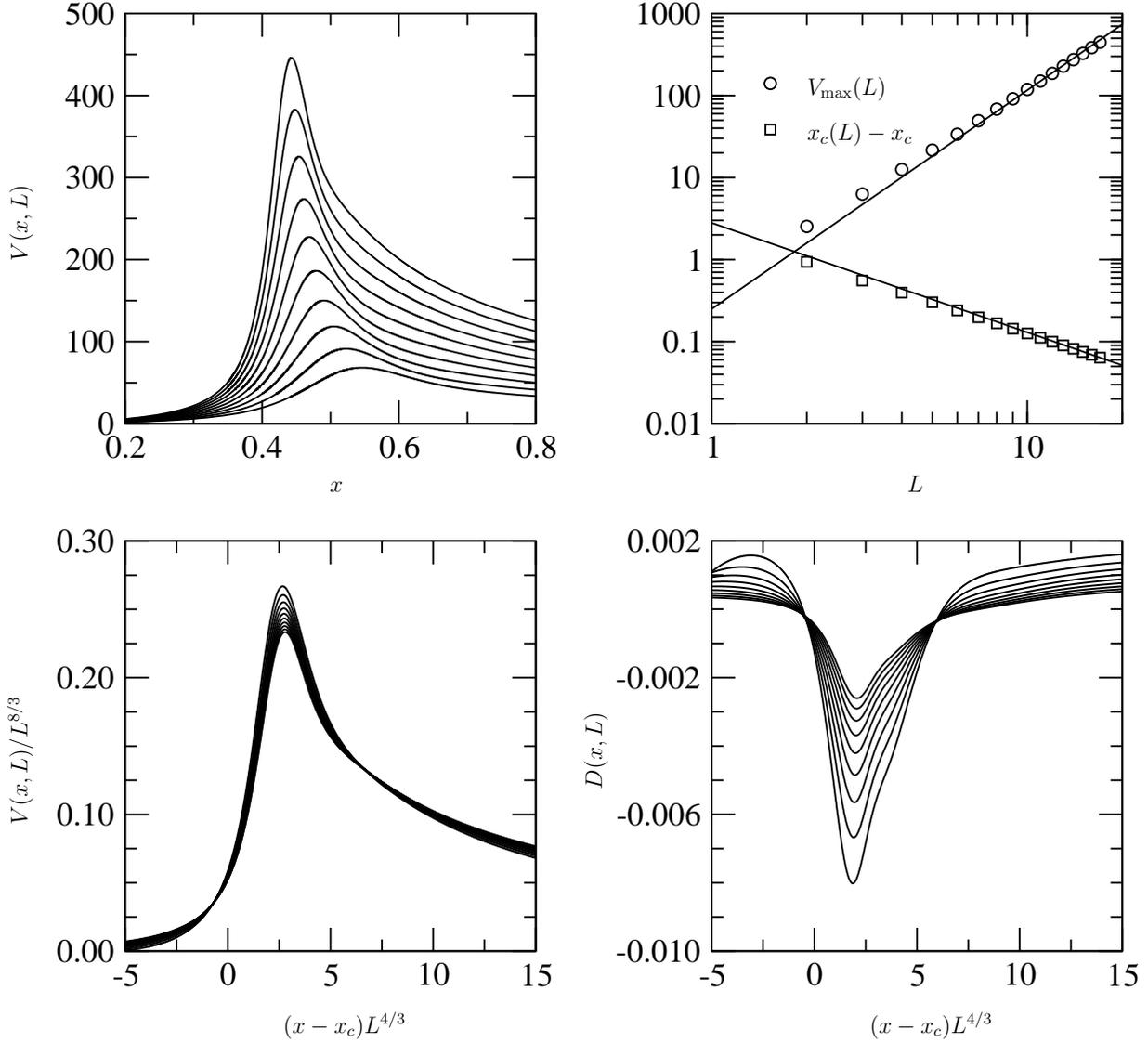}
\caption{\label{fig:fluc}
The fluctuations $V(x,L)$ as a function of the fugacity $x$ (top left panel).
$x_c(L)-x_c$ and $V_{\rm max}$ vs. $L$ (top right panel).
$V(x,L)/L^{8/3}$ (bottom left panel) and $D(x,L)$ (bottom right panel)
vs. the scaling variable $(x-x_c)L^{4/3}$.
}
\end{figure}

Whittington and Guttmann \cite{WG} and later Burkhardt and Guim \cite{BG91}
 studied the behaviour of the mean number of steps
in a path on an $L \times L$ lattice
\begin{equation}
\< n(x,L) \> = \frac{\sum_n n c_n(L) x^n}{\sum_n c_n(L) x^n}
\end{equation}
as well as the fluctuations of this quantity
\begin{equation}
V(x,L) = \frac{\sum_n n^2 c_n(L) x^n}{\sum_n c_n(L) x^n}-\< n(x,L) \>^2
\end{equation}
which is a kind of heat capacity. As discussed above, a phase transition takes place as
one varies the fugacity $x$ associated with the walk length.
At a critical value $x_c$, the  average walk length
of a path on an $L \times L$ lattice changes
from $\Theta(L)$ to $\Theta(L^2).$ In \cite{WG} the critical fugacity was proved
to satisfy $1/\mu \le x_c \le \mu_H$, where $\mu_H$ is
the growth constant for Hamiltonian
SAW on the square lattice, and on the basis of numerical studies
conjectured to be $x_c = 1/\mu$ exactly.  In \cite{M} the conjecture was proved.
Here we also study the
behaviour at $x=x_c$ and find that $\< n(x,L) \> = \Theta( L^{1/\nu})$ where the
numerical evidence is consistent with $\nu = 3/4$. Similar conclusions were reached earlier
in \cite{BG91}.
For any given value of $L$ the fluctuation $V(x,L)$ is observed to have a single maximum
located at $x_c(L)$ (see top left panel of Figure~\ref{fig:fluc}).
We study in detail the behaviour of $ V(x,L)$, which we  expect to obey a
standard finite-size scaling Ansatz
\begin{equation}
V(x,L) \sim L^{2/\nu}\tilde{V}((x-x_c)L^{1/\nu}),
\end{equation}
(which is equivalent to (\ref{BG}) of \cite{BG91})
where $\tilde{V}(y)$ is a scaling function. From this it follows that
the position and the height of the peak in $V(x,L)$ scale as
$x_c(L)-x_c \sim L^{-1/\nu}$ and $V_{\rm max}(L) \sim L^{2/\nu}$.

\begin{table}[h]
\centering
\caption{\label{tab:flucval} The mean-length of walks crossing an $L\times L$ square
at the critical fugacity $x=x_c$, the position, $x_c(L)-x_c$, and height, $V_{\rm max}(L)$,
of the peak in the fluctuations $V(x,L)$.}
\begin{tabular}{llll}  \hline  \hline
$L$ & $\< n(x_c,L) \> $ & $x_c(L)-x_c$ & $V_{\rm max}(L)$  \\ \hline
1   &   2  & & \\
2   &   4.1230827138  &  0.9370217352   &      2.5358983849 \\
3   &   6.3491078353  &  0.5554687338   &      6.2850743202 \\
4   &   8.6519365910  &  0.3963960508   &     12.5671289312 \\
5   &  11.0129773423  &  0.3016714640   &     21.6246676036 \\
6   &  13.4187561852  &  0.2403448999   &     33.7507328831 \\
7   &  15.8593480600  &  0.1979673072   &     49.2268220069 \\
8   &  18.3273545355  &  0.1671981710   &     68.3294309970 \\
9   &  20.8171976528  &  0.1439801106   &     91.3288825240 \\
10  &  23.3246243077  &  0.1259158112   &    118.4887185709 \\
11  &  25.8463556412  &  0.1115091953   &    150.0657089122 \\
12  &  28.3798369044  &  0.0997832765   &    186.3101460060 \\
13  &  30.9230572826  &  0.0900753740   &    227.4662469752 \\
14  &  33.4744187854  &  0.0819213689   &    273.7725788463 \\
15  &  36.0326398605  &  0.0749872153   &    325.4624696518 \\
16  &  38.5966838209  &  0.0690267737   &    382.7643901657 \\
17  &  41.1657051788  &  0.0638549420   &    445.9023015941 \\
 \hline  \hline
\end{tabular}
\end{table}

In table~\ref{tab:flucval} we have listed the numerical values of
the mean-length at $x_c$ and the position and height of the
maximum of the fluctuations. We analyse this data by forming the
associated generating functions, $N(z)= \sum_L \< n(x,L) \> z^L$
etc., and using differential approximants. Given the expected
asymptotic behaviour of these quantities the generating functions
should have a singularity at $z_c=1$ with critical exponents
$-1/\nu-1$ (average length at $x_c$),  $1/\nu-1$ (position of the
peak), and $-2/\nu-1$ (height of the peak). In
table~\ref{tab:flucana} we list the results from an analysis of
the generating functions using second order differential
approximants. The estimates for the exponents are not very
accurate (which is not surprising given the short length of the
series) but are fully consistent with $\nu = 3/4$.

Finally, in Figure~\ref{fig:fluc} we perform a more detailed
analysis to confirm the conjectured scaling form for $V(x,L)$. In
the top left panel we have simply plotted $V(x,L)$ as a function
of the fugacity $x$ to confirm the single peak behaviour. In the
top right panel we have plotted $x_c(L)$ and $V_{\rm max}$ vs. $L$
in a log-log plot, thus confirming that these quantities grows as
a power-law with $L$ (the straight lines, drawn as a guide to the
eye,  have slopes $-1/\nu=-4/3$ and $2/\nu=8/3$, respectively). In
the bottom panels we check numerically the scaling Ansatz for
$V(x,L)$. In the left panel we plot $V(x,L)/L^{8/3}$ vs. the
scaling variable $(x-x_c)L^{4/3}$ obtaining a reasonable scaling
collapse. A better idea of the quality of the scaling collapse can
be gauged from the plot in the bottom right panel. Here we plot
the difference between consecutive scaling plots from the left
panel. More precisely we plot
$D(x,L)=V(x,L)/L^{8/3}-V(x',L-1)/(L-1)^{8/3}$ vs.
$(x-x_c)L^{4/3}$, where $x'$ is chosen so that the scaled
variables coincide, e.g., $(x-x_c)L^{4/3}=(x'-x_c)(L-1)^{4/3}$.

\begin{table}
\centering \caption{\label{tab:flucana} Estimates for $z_c$ and
the critical exponents obtained from second order differential
approximants to the generating functions in
Table~\protect{\ref{tab:flucval}}. $K$ is the degree of the
inhomogeneous polynomial of the differential approximant.}
\begin{tabular}{lllllll}  \hline  \hline
$K$ & \multicolumn{2}{c}{$\< n(x_c,L) \>$}
   & \multicolumn{2}{c}{$x_c(L)-x_c$}
   & \multicolumn{2}{c}{$V_{\rm max}(L)$}  \\ \hline
   &  $z_c$ & $-1/\nu-1$ &  $z_c$ & $1/\nu-1$ &  $z_c$ & $-2/\nu-1$ \\ \hline
0 & 0.9999823(13)& -2.32985(17)& 1.00017(11) & 0.3147(96)& 0.999998(20) & -3.6620(21) \\
1 & 0.999983(10) & -2.3299(12) & 1.000114(23)& 0.3196(16)& 0.9999900(41)& -3.66134(34) \\
2 & 0.9999818(79)& -2.32973(99)& 1.000124(15)& 0.3185(16)& 0.999982(10) & -3.6606(10)\\
3 & 0.9999789(88)& -2.3293(11) & 1.00013(10) & 0.3183(81)& 0.999975(17) & -3.6598(18)\\
4 & 0.9999773(76)& -2.32915(93)& 1.000084(45)& 0.3215(47)& 0.999979(11) & -3.6603(11)\\
5 & 0.9999786(70)& -2.32930(80)& 1.000136(75)& 0.3171(61)& 0.9999850(69)& -3.66081(65)\\
 \hline  \hline
\end{tabular}
\end{table}

\section{Walks crossing the square and hitting the centre}

In \cite{K76} Knuth also considered the problem of self-avoiding walks
crossing
the square {\em and} passing through the centre $(L/2,L/2)$ of the grid
(with
$L$ being even). Denote the number of such walks by  $c(L).$
Then a straightforward variant of the method of proof used
in Section \ref{sec:altbounds} can be applied to prove that
$$\lim_{L \to \infty} c(L)^{1/L^2} = \lambda^2.$$
Knuth used Monte Carlo simulations to estimate the
fraction of
paths hitting the centre point and found for $L=10$ that $81 \pm 10$
percent
of all paths do hit the centre. He then went on to say that ``perhaps
nobody
will ever know the true answer.'' Naturally we cannot let Knuth's
challenge
go unanswered. It is very simple to modify the transfer-matrix algorithm
to
ensure that all paths pass through a given vertex. We just make sure
that when
we do the updating at the given vertex the input state $'00'$ (no
occupied incoming edges)
has only one output state $'12'$, while the output $'00'$ (no outgoing
occupied edges)
is disallowed at this vertex. We can thus answer Knuth's query and
state for all
to know that for $L=10$ a fraction
$1243982213040307428318660/1568758030464750013214100=
0.792972 \ldots$ of all paths pass through the centre. In
Table~\ref{tab:centre}
we have listed the number of paths passing through the centre for
$L\leq 18$.

\begin{table}
\centering
\caption{\label{tab:centre} The total number of walks crossing an $L \times L$
 square and passing through the centre $(L/2,L/2)$, $c(L)$ and the ratio $c(L)/C(L).$ }
\renewcommand{\arraystretch}{1.05}
\scriptsize
\begin{tabular}{rll}  \hline  \hline \\
   $L$  &  $c(L)$ & $c(L)/C(L)$ \\
\hline
  2 & 10
    & $0.833333\ldots$ \\
  4 & 7056
    & $0.828947\ldots$ \\
  6 & 462755440 & $0.803701 \ldots$\\
  8 & 2593165016903538
    & $ 0.793842\ldots$ \\
10 & 1243982213040307428318660
    & $0.792972 \ldots$  \\
12 & 51166088445891978924432033203830714
    & $0.792927 \ldots$  \\
14 & 180349587397776823066172713933745722978533730900
    & $0.792920 \ldots$  \\
16 & 54508896286415931462305055600895616388822171335171594099162334
    & $0.792909 \ldots$  \\
18 &
1413040380714086952244299343879218154884335669707058802937825791571640010167156
    & $0.792901 \ldots$  \\
\hline \hline
\end{tabular}
\end{table}

The fact that $C(L)/c(L)$ appears to be going to a constant implies that not only
is the asymptotically dominant behaviour of both $C(L)$ and $c(L)$ the same, but so
must the sub-dominant behaviour. We note the useful mnemonic that the ratio appears
close to $\sqrt{\pi/5} = 0.79266...,$ though we have no idea how to prove or disprove
that this is the correct
value.

\section{Hamiltonian walks  \label{sec:hamilton}}

Hamiltonian walks can only exist on $2L \times 2L$ lattices. For lattices with an odd
number of edges, one site must be missed. A Hamiltonian walk is of length $4L(L+1)$ on
a $2L \times 2L$ lattice. The number of such walks grows as $\tau^{4L^2},$ where we
find $\tau \approx 1.472$ based on exact enumeration up to $17 \times 17$ lattices.
In \cite{J98} Jacobsen and Kondev gave a field-theoretical estimate of the growth constant 
for Hamiltonian SAW on the square lattice as $1.472801 \pm 0.00001$. These were walks confined 
to a square geometry, but not restricted as to starting and end-points as are those we consider 
here. Nevertheless, it seems likely that we are estimating the same quantity, so our results
can be seen as providing support for the view that the field theory is estimating precisely
the same quantity as our enumerations. That is to say, this appears to be precisely the same 
as the corresponding result for Hamiltonian walks on an $L \times L$ lattice, in the large $L$ 
limit. These estimates are about $20\%$ less than $\lambda,$ the growth
constant for all paths. In \cite{AH} it is proved that $2^{1/3} \le \tau \le 12^{1/4}.$
Numerically this evaluates to $1.260 \le \tau \le 1.861.$

We can improve on these bounds as follows: we define  cow-patch walks to be
Hamiltonian if every vertex of the square not belonging to the
border of the square belongs to one of the SAWs of the cow-patch. Then the upper bounds given
above translate verbatim into upper bounds for $\tau,$ while lower bounds are given
by Hamiltonian traversing paths and eqn. (\ref{lower2}). In this way we find $1.429 < \tau < 1.530.$
As we have shown above that $1.6284 < \lambda,$ this proves that $\tau < \lambda.$

The number of Hamiltonian paths $H_L$ for $L$ even, and paths that
visit all but one site, for $L$ odd, are given in Table
\ref{tab:man}. The number of Hamiltonian cow-patch paths $HP_L$
for $L$ even, and cow-patch paths that visit all but one site, for
$L$ odd, are given in Table \ref{tab:manc}. The number of
Hamiltonian transverse paths $HT_L$ for $L$ even, and transverse
paths that visit all but one site, for $L$ odd, are given in Table
\ref{tab:mans}.

\begin{table}
\centering
\caption{\label{tab:man} The number of Hamiltonian paths.}
\renewcommand{\arraystretch}{1.05}
\scriptsize
\begin{tabular}{rl}  \hline  \hline \\
  $L$  &  $H_L$ \\
\hline
 1 &  2  \\
 2 &  2  \\
 3 &  32  \\
 4 &  104 \\
 5 &  10180 \\
 6 &  111712 \\
 7 &  67590888 \\
 8 &  2688307514 \\
 9 &  9628765945000 \\
10 &  1445778936756068 \\
11 &  29725924602729604016 \\
12 &  17337631013706758184626 \\
13 &  1998903003325610328086958408 \\
14 &  4628650743368437273677525554148 \\
15 &  2937440223891635053435045277805847436 \\
16 &  27478778338807945303765092195103685118924 \\
17 &  94555056448262478314997568263027383699860223148 \\
\hline \hline
\end{tabular}
\end{table}
\begin{table}
\centering
\caption{\label{tab:manc} The number of Hamiltonian cow-patch paths.}
\renewcommand{\arraystretch}{1.05}
\scriptsize
\begin{tabular}{rl}  \hline  \hline \\
  $L$  &  $HP_L$ \\
\hline
  2 &6 \\
  3 &81 \\
  4 &2420 \\
  5 &158487 \\
  6 &22668546 \\
  7 &7067228903 \\
  8 &4796951277784 \\
  9 &7083189530689311 \\
10 &22740544515287098346 \\
11 &158673902903632923216807 \\
12 &2405521769596577026409223804 \\
13 &79215226453280152797069512845071 \\
14 &5665275864000731097175367200188234758 \\
15 &879791999732650875090633720304683597787867 \\
16 &296640712696590626976673730832416228749213171388 \\
17 &217134088450048497810206709994144694071029172119163041 \\
18 &345011492148033546292595301223727273934239259467419472922686 \\

\hline
\hline \hline
\end{tabular}
\end{table}
\begin{table}
\centering
\caption{\label{tab:mans} The number of Hamiltonian traversing paths.}
\renewcommand{\arraystretch}{1.05}
\scriptsize
\begin{tabular}{rl}  \hline  \hline \\
  $L$  &  $HT_L$ \\
  1 &2 \\
  2 &8 \\
  3 &34 \\
  4 &650 \\
  5 &12014 \\
  6 &1016492 \\
  7 &83761994 \\
  8 &32647369000 \\
  9 &12227920752840 \\
10 &22181389298814376 \\
11 &38166266554504010420 \\
12 &323646210116765453608746 \\
13 &2574827340090912815899810042 \\
14 &102299512403818451392332665527950 \\
15 &3778748215131699995997836850757543682 \\
16 &704314728645701361948084580318587261484806 \\
17 &121135616205759617794904559766506890558675949856 \\
18 &106005756542854454380006180528618254764945283647525384 \\

\hline \hline
\end{tabular}
\end{table}

\section*{E-mail or WWW retrieval of series}

The series for the problems studied in this paper are available
by request from
I.Jensen@ms.unimelb.edu.au or via the world wide web
http://www.ms.unimelb.edu.au/\~{ }iwan/ by following the relevant links.

\section*{Acknowledgments}
We would like to thank Stu Whittington for helpful comments on the manuscript.
MBM  was partially supported by the European Commission's IHRP
  Programme, grant HPRN-CT-2001-00272, ``Algebraic Combinatorics in
  Europe''.
Two of us (AJG and IJ) have been supported by grants from
the Australian Research Council.
The calculations presented in this paper were performed on facilities
provided by the Australian Partnership for Advanced Computing (APAC) and
the Victorian Partnership for Advanced Computing (VPAC).

 \end{document}